\documentclass[a4paper,10pt,english,twocolumn,amsmath,amssymb,showpacs,notitlepage,nofootinbib,longbibliography]{revtex4-1}
\usepackage{graphicx}
\usepackage{mathrsfs}
\usepackage[unicode=true,pdfusetitle,
bookmarks=true,bookmarksnumbered=false,bookmarksopen=false,
breaklinks=false,pdfborder={0 0 1},colorlinks=false]{hyperref}
\hypersetup{colorlinks,citecolor=blue,linkcolor=blue}

\begin{document}

	\title{Understanding the negative binomial multiplicity fluctuations in relativistic heavy ion collisions}

\author{Hao-jie Xu}

\email{haojiexu@pku.edu.cn}

\affiliation{Department of Physics and State Key Laboratory of Nuclear Physics
and Technology, Peking University, Beijing 100871, China}

\date{\today}

\begin{abstract} 
	By deriving a general expression for multiplicity distribution (a conditional probability distribution) in statistical
	model, we demonstrate the mismatches between experimental measurements and previous theoretical calculations on multiplicity fluctuations. 
	From the corrected formula, we develop an improved baseline measure for multiplicity distribution under Poisson approximation  in
	statistical model to replace the traditional Poisson expectations. We find that the ratio of the mean multiplicity to the 
	corresponding reference multiplicity are crucial to systemically explaining the measured scale variances of total charge distributions 
	in different experiments, as well as understanding the centrality resolution effect observed in experiment. The improved statistical expectations, albeit simple, work well in 
	describing the negative binomial multiplicity distribution  measured in experiments, e.g. the cumulants (cumulant products) of total
	(net) electric charge  distributions.
\end{abstract}


\maketitle

\section{Introduction} 

Event-by-event multiplicity fluctuations are expected to provide us crucial informations
about the hot and dense Quantum chromodynamics (QCD) matter created in heavy ion collision (HIC)~\cite{Stephanov:1999zu,Jeon:1999gr,Stephanov:2008qz,Bazavov:2012vg,Gupta:2011wh}.
In experiment~\cite{Aggarwal:2010wy,Adamczyk:2013dal,Adler:2007fj,Adare:2008ns,Adamczyk:2014fia,Tang:2014ama,Adare:2015aqk}, 
the multiplicity distribution of total (net-conserved) charges 
published by STAR and PHENIX Collaboration were calculated using particles with 
specific kinematic cuts (denoted as sub-event $B$), and the
centrality cuts were made using  particles with some other acceptance windows 
(denoted as sub-event $A$).  To avoid auto correlation,
these two sub-events have been separated by different pseudorapidity
intervals or particle species. For example, in the net-charges case~\cite{Adamczyk:2014fia},
the kinematic cut for the centrality-definition
particles in sub-event $A$ is $1.0>|\eta|>0.5$ and for the moment-analysis particles
in sub-event $B$ is $|\eta|<0.5$, where $\eta$ is pseudorapidity. In this
work, we always use $q$ to represent the multiplicity in sub-event $B$ for the
study of multiplicity distribution, and use $k$ to represent the multiplicity
in sub-event $A$ for the centrality definition. The latter $k$ is also called
reference multiplicity.  It is observed in experiments~\cite{Adler:2007fj,Adare:2008ns,Adamczyk:2014fia,Tang:2014ama,Adare:2015aqk} 
that the (total, positive, negative, net) charge 
distribution can be well described  by the negative binomial distribution (NBD),
\begin{equation}
		\mathrm{NBD}(q;p,r)\equiv \frac{(q+r-1)!}{q!(r-1)!}p^{q}(1-p)^{r},
		\label{eq:NBD}
\end{equation} 
where $p$ ($0<p<1$) is the success probability in each trial, and $q$ ($r$) is 
the number of success (failure). 

Due to its success in describing the ratios of particle multiplicities data in a broad 
energy range of relativistic heavy ion collisions (see e.g. \cite{Andronic:2005yp} and the references therein),
the statistical model and its variations has been regarded as one of the basic tools in studying the
baseline prediction for the data on multiplicity fluctuations~\cite{Cleymans:2004iu,Begun:2004gs,Begun:2006uu,Garg:2013ata,Karsch:2010ck,Alba:2014eba,Fu:2013gga,Bhattacharyya:2015zka,BraunMunzinger:2011dn,Karsch:2015zna,Bzdak:2012an}.
For the mathematical convenience, the Poisson distribution, which can be obtained from grand canonical ensemble (GCE) with Boltzmann statistics~\cite{Cleymans:2004iu,Begun:2004gs,Karsch:2010ck} have been frequently used in HIC as one basic baseline measure for multiplicity fluctuations~\cite{Aggarwal:2010wy,Adamczyk:2013dal,Adamczyk:2014fia}. 
To understand the deviations of data from Poisson distributions, there are many effects have been studied in statistical models, e.g. finite volume effect, quantum effect,
 experimental acceptance, as well as the resonance decays which were once considered as one of the major contributions to the deviations. 
Despite many improvements of statistical models~\cite{Cleymans:2004iu,Begun:2004gs,Begun:2006uu,Garg:2013ata,Karsch:2010ck,Alba:2014eba,Fu:2013gga,Bhattacharyya:2015zka,BraunMunzinger:2011dn,Karsch:2015zna,Bzdak:2012an},
however, there are still  
difficulties in their systemically describing the data on negative binomial multiplicity distributions. For example, the measured scale variation of total charge distributions are very different in different centralities and different experiments~\cite{Alt:2006jr,Adare:2008ns,Tang:2014ama,Mukherjee:2016hrj}. 
This implies that some external effects~\cite{Gorenstein:2008th,Gorenstein:2011vq,Gorenstein:2013nea,Skokov:2012ds,Stodolsky:1995ds,Wilk:2009nn,Biro:2014yoa},
unrelated to the critical phenomenon, should be 
included. Recently, the effect of volume fluctuations on cumulants of multiplicity distributions have been studied by Skokov and his collaborations~\cite{Skokov:2012ds}.

Unfortunately, previous theoretical studies are only focus on the probability distribution $P_{B}(q)$(without volume fluctuations)
or $\mathscr{P}_{B}(q)$(with volume fluctuations), but overlook the effect of probability conditions from sub-event A, 
here we postpone the definitions of $P_{B}(q)$ and $\mathscr{P}_{B}(q)$ to the next section (see Eq.~\ref{eq:Bmultiplicity}). We will show that neither $P_{B}(q)$ nor $\mathscr{P}_{B}(q)$  is
the correct formula of probability distribution in describing the experimental measurements on multiplicity fluctuations.
Clarifying the mismatches between the experiments and the previous theoretical calculations 
on multiplicity distributions and then understanding the negative binomial multiplicity distributions of electric charges 
are the main motivation of this work.

The main observation of this work is that: after including the distribution of principal thermodynamic variables (PTVs) 
in statistical model (e.g.,distribution of volume, the dominated effect in HIC), 
the sub-event $A$ and $B$ corresponding to  the method used in experiments are correlated to 
each other in event-by-event analysis, and, as far as we know, 
this feature have not been taken seriously in previous studies. These correlations make the 
measured multiplicity distribution becomes a conditional probability distribution (Eq.~(\ref{eq:generaleq})), 
instead of the traditional probability distribution (Eq.~(\ref{eq:Bmultiplicity}))
discussed in previous studies~\cite{Cleymans:2004iu,Begun:2004gs,Begun:2006uu,Garg:2013ata,Karsch:2010ck,Alba:2014eba,Fu:2013gga,Bhattacharyya:2015zka,BraunMunzinger:2011dn,Karsch:2015zna,Bzdak:2012an,Gorenstein:2008th,Gorenstein:2011vq,Gorenstein:2013nea,Skokov:2012ds}.
We develope an improved baseline measure for multiplicity distribution under Poisson approximation in statistical model with the corrected probability distributions. The improved statistical expectations, albeit simple, work well in describing the negative binomial multiplicity distribution  measured in experiments, e.g.,
\begin{itemize}
	\item The relations among the scale variances of positive, negative and total charge distributions reported by the NA49 Collaboration~\cite{Alt:2006jr} and the PHENIX Collaboration~\cite{Adare:2008ns}.
	\item The variances of total charge distributions at $\sqrt{s_{NN}}=27$ GeV reported by the STAR Collaboration~\cite{Tang:2014ama}.
	\item The sensitivity  of NBD parameters on the transverse momentum range of momentum-analysis particles reported by the PHENIX Collaboration~\cite{Adare:2008ns}.
	\item The NBD baselines used for the cumulant products of net-charge distributions reported by the STAR Collaboration~\cite{Sarkar:2014fja}. 
	\item The differences between the  cumulants of net-charges and net-kaons distributions reported by the STAR Collaboration~\cite{Adamczyk:2014fia,Sarkar:2014fja}.
  \item The centrality resolution effect observed in experiment~\cite{Luo:2013bmi}. 
\end{itemize}
The results indicate that the probability conditions from sub-event A play  crucial roles to explain the negative binomial multiplicity 
distributions of (net) electric charges measured in sub-event B.

The paper is organized as follows. In Sec.~\ref{derivation}, 
we will demonstrate the mismatches between experimental measurements and previous theoretical calculations, 
by deriving a general formula for the multiplicity fluctuation 
corresponding to the method used in experiment~\cite{Adler:2007fj,Adare:2008ns,Aggarwal:2010wy,Adamczyk:2013dal,Adamczyk:2014fia,Tang:2014ama,Adare:2015aqk}.
In Sec.~\ref{baseline}, under Poisson approximation, we 
will show how to calculate the improved statistical baseline measure for 
higher order cumulants of multiplicity distributions. We will also give  
approximate formula for higher order cumulants which can explain most of  
experimental observables related to multiplicity fluctuations such as the scale 
variance, the centrality resolution effect, et. al. We will give a summary in the final 
section.

\section{General derivation} 
\label{derivation}

In this section, we derive a general expression for the multiplicity
distribution, related to recent experiments at RHIC~\cite{Adler:2007fj,Adare:2008ns,Aggarwal:2010wy,Adamczyk:2013dal,Adamczyk:2014fia,Tang:2014ama,Adare:2015aqk}. 
To avoid centrality bin width effect in experiment, the cumulant calculations are restricted in a
fine bin of centrality (a reference multiplicity bin is the finest centrality bin)~\cite{Tang:2014ama,Luo:2013bmi},
the bin width depend on the statistics.  In
this work, we calculate the cumulants of multiplicity distribution as function
of reference multiplicity, the relation between the results in reference
multiplicity bin and in centrality bin are obvious.

In a specific statistical ensemble (SSE), the probability distribution of multiplicity $X$ is defined as
$P_{E}(X;\mathbf{\Omega})$,  where $\mathbf{\Omega}$ represents a set of PTVs
(e.g.,for GCE, $\mathbf{\Omega}=(T,V,\mu)$). After employing the distribution of PTVs
$F(\mathbf{\Omega})$, which was caused by the collisional geometry in HIC, we obtain the multiplicity distribution in statistical model~\cite{Gorenstein:2008th,Skokov:2012ds}
\begin{equation}
	\mathscr{P}(X)=\int d\mathbf{\Omega}F(\mathbf{\Omega})P_{E}(X;\mathbf{\Omega}).
	\label{eq:genmultiplicity}
\end{equation}
On experimental side, $\mathscr{P}(X)$\footnote{We always
use $P$ to represent the probability distribution in a SSE, and use $\mathscr{P}$ to
represent the probability distribution measured in experiment.} stand for  the multiplicity
distribution measured in a specific acceptance windows (e.g.,rapidity,
pseudorapidity, transverse momentum, particle species, et.al.). It can be
used for centrality definition or for moment analysis. Meanwhile,
Eq.(\ref{eq:genmultiplicity}) can be also regarded as the general formula
of $\alpha$-ensemble discussed in Ref.~\cite{Gorenstein:2008th}.

From Eq.(\ref{eq:genmultiplicity}), the distribution of reference multiplicity 
$k$ and the distribution of multiplicity $q$ can be written as
\begin{eqnarray}
	\mathscr{P}_{A}(k) &=& \int d\mathbf{\Omega}F(\mathbf{\Omega})P_{A}(k;\mathbf{\Omega}),\label{eq:genrefmultiplicity} \\
	\mathscr{P}_{B}(q) &=& \int d\mathbf{\Omega}F(\mathbf{\Omega})P_{B}(q;\mathbf{\Omega}),\label{eq:Bmultiplicity}
\end{eqnarray}
where $P_{A}(k;\mathbf{\Omega})$ and $P_{B}(q;\mathbf{\Omega})$
stand for multiplicity distribution in a SSE with specific acceptance cuts
for sub-event $A$ and sub-event $B$, respectively. 

It is worth noting that, although $\mathscr{P}_{A}(k)$ can been regarded as
distribution of reference multiplicity measured in experiment, neither
$P_{B}(k;\mathbf{\Omega})$ nor $\mathscr{P}_{B}(k)$ can be used to represent
the experiment measurements~\cite{Aggarwal:2010wy,Adamczyk:2013dal,Adamczyk:2014fia,Tang:2014ama,Adare:2015aqk}.
This is because the multiplicity 
distribution of moment-analysis particles measured in experiment is a 
\textbf{conditional probability distribution}.  Briefly stated,
condition refers to the notion that the calculations of cumulants 
are restricted in a specific centrality (reference multiplicity)
bin. 
We note that $P_{B}(k;\mathbf{\Omega})$ and $\mathscr{P}_{B}(k)$ are independent of the definition of reference multiplicity,  
and they have been widely discussed in previous studies~\cite{Cleymans:2004iu,Begun:2004gs,Begun:2006uu,Garg:2013ata,Karsch:2010ck,Alba:2014eba,Fu:2013gga,Bhattacharyya:2015zka,BraunMunzinger:2011dn,Karsch:2015zna,Bzdak:2012an,Gorenstein:2008th,Gorenstein:2011vq,Gorenstein:2013nea,Skokov:2012ds}. Unfortunately, both of them are not the correct formula for the multiplicity distributions measured in experiment.

The conditional probability distribution for multiplicity $q$
in  given reference multiplicity bin $k$ reads, 
\begin{equation}
	\mathscr{P}_{B|A}(q|k)=\frac{\mathscr{P}_{A\cap B}(q,k)}{\mathscr{P}_{A}(k)}.
\label{eq:md} 
\end{equation} 
where
\begin{equation}
	\mathscr{P}_{A\cap B}(q,k)=\int d\mathbf{\Omega}F(\mathbf{\Omega})P_{A\cap B}(q,k;\mathbf{\Omega})
	\label{eq:twomultiplicity}
\end{equation} 
and  $P_{A\cap B}(q,k;\mathbf{\Omega})$ is a joint
probability distribution for sub-event $A$ and $B$ in a SSE.  With some
experimental techniques, the two sub-events are expected to be independent of
each other. In this case, we have 
\begin{equation} 
	P_{A\cap B}(q,k;\mathbf{\Omega}) = P_{B}(q;\mathbf{\Omega})P_{A}(k;\mathbf{\Omega}).
\end{equation} 
In this work, we focus on such independent approximation.
We note that, due to dynamic evolution and the correlation between different
particle species in HIC, the independent approximation might be contaminated.

With independent approximation, Eq.(\ref{eq:md}) can be written as
\begin{equation} 
	\mathscr{P}_{B|A}(q|k)=\frac{\int d\mathbf{\Omega}F(\mathbf{\Omega})P_{B}(q;\mathbf{\Omega})P_{A}(k;\mathbf{\Omega})}{\mathscr{P}_{A}(k)}.
\label{eq:generaleq} 
\end{equation}
Consequently, we derive a general expression in statistical model for arbitrary statistical
ensemble and arbitrary distribution of PTVs, related to recent
data~\cite{Aggarwal:2010wy,Adamczyk:2013dal,Adamczyk:2014fia,Tang:2014ama}
on multiplicity distributions.  For a specific calculation,
the informations of $P_{A}(k;\mathbf{\Omega})$,
$P_{B}(q;\mathbf{\Omega})$, as well as $F(\mathbf{\Omega})$ are required. 

Due to $P_{A}(k;\mathbf{\Omega})$ and $F(\mathbf{\Omega})$ appeared in both
Eq.(\ref{eq:genrefmultiplicity}) and Eq.(\ref{eq:generaleq}), the connection
between the distribution of reference multiplicity $\mathscr{P}_{A}(k)$ and
multiplicity distribution of moment-analysis particles $\mathscr{P}_{B|A}(q|k)$
has been established. In the next section, we will show that this connection
is crucial to explain the centrality resolution effect measured in experiment~\cite{Luo:2013bmi}.

\section{Applications: statistical expectations under Poisson approximation} 
\label{baseline}
In this section, we calculate the improved
baseline measure of cumulants of multiplicity fluctuations under a simple approximation: 
$P_{A}(k;\mathbf{\Omega})$ and $P_{B}(q;\mathbf{\Omega})$, the distributions
in a SSE, can be regarded as Poisson distributions. 
In a SSE~\cite{Becattini:1995xt,Stephanov:1999zu,Cleymans:2004iu,BraunMunzinger:2011ta,Bhattacharyya:2015zka}, there are
many other effects that make the distribution deviates
from Poisson distribution,
e.g., finite volume effect, quantum effect,
resonance decays, experimental acceptance, et.al, which can be a topic for our future study. 

The outline of the present section is as follows. In Sec.~\ref{sec:refmul}, we calculate the cumulants of 
$\mathscr{P}_{A}(k)$ and $\mathscr{P}_{B|A}(q|k)$ under Poisson approximation.
With the help of the data of reference multiplicity $\mathscr{P}_{A}(k)$ and
mean value distribution $\mathscr{M}(k)$ measured in experiment, we demonstrate
how to obtain the higher order cumulants of multiplicity distribution in the 
improved statistical model. The calculations are directly applied to the 
net-conserved charges case in Sec.~\ref{sec:netcharge}.
In Sec.~\ref{sec:approx}, we calculate the approximate solutions of 
these higher order cumulants which can explain most of the
experiment observables. Finally, in Sec.~\ref{sec:discussion}, a short
discussion is given to highlight some of the difficulties in the improved
statistical baseline measure.

\subsection{Improved statistical baseline measure} 
\label{sec:refmul} 
In this section, we consider the discussion of one PTV, e.g., the system
volume as the dominated effect in HIC. With Poisson approximation 
\begin{equation}
P_{A}(k;\lambda)=\frac{\lambda^{k}e^{-\lambda}}{k!}
\label{eq:multiplicity0}
\end{equation}
for sub-event $A$, where the Poisson parameter
$\lambda\equiv\lambda(\mathbf{\Omega})$ is determined by 
$\Omega$ and acceptance cuts, the distribution of reference
multiplicity $\mathscr{P}_{A}(k)$ in Eq.(\ref{eq:genrefmultiplicity}) can be
written as, 
\begin{eqnarray} 
	\mathscr{P}_{A}(k)&=&\int d\mathbf{\Omega}F(\mathbf{\Omega})\frac{\lambda^{k}e^{-\lambda}}{k!}\nonumber\\
							&=&\int d\lambda f(\lambda)\frac{\lambda^{k}e^{-\lambda}}{k!}, 
	\label{eq:multiplicity}
\end{eqnarray} 
where $f(\lambda)$ is the normalized distribution of Poisson
parameter.  The scale variance of $\mathscr{P}_{A}(k)$ reads
\begin{equation}
	\omega_{A}\equiv\frac{\sigma^{2}_{A}}{M_{A}}=1+\frac{\int d\lambda f(\lambda)(\lambda-M_{A})^{2}}{M_{A}},
	\label{eq:scalevariance}
\end{equation}
where $M_{A} = \int d\lambda f(\lambda)\lambda$ and
$\sigma^{2}_{A}$ are the mean value and variance of $\mathscr{P}_{A}(k)$ . The
most significant feature of Eq. (\ref{eq:scalevariance}) is that we obtain
$\omega_{A}>1$ except one special case $f(\lambda)=\delta(M)$~\footnote{This feature might be interesting in elementary nucleon-nucleon
collisions. Because we notice that in this case, $\mathscr{P}(k)$ have been solely used to
calculate the corresponding cumulants, and the results show a typical NBD
feature: $\omega>1$~\cite{Arneodo:1987qy,Adamus:1987ea,Ansorge:1988kn,Aamodt:2010ft,Khachatryan:2010nk}.}.

\label{sec:expectations}
Using Poisson approximation for both sub-event $A$ and sub-event $B$, we
obtain the conditional probability distribution from Eq.(\ref{eq:generaleq}) as
\begin{eqnarray} 
	\mathscr{P}_{B|A}(q|k) &=&\frac{1}{\mathscr{P}_{A}(k)}\int d\mathbf{\Omega} F(\mathbf{\Omega})\frac{\lambda^{k}e^{-\lambda}}{k!} \frac{\mu^{q}e^{-\mu}}{q!}\nonumber \\
								&=&\mathscr{N}(k)\int d\lambda f(\lambda)\frac{\lambda^{k}e^{-\lambda}}{k!}\frac{\mu^{q}e^{-\mu}}{q!},
	\label{eq:distribution}
\end{eqnarray}
where $\lambda$, $\mu= \mu(\mathbf{\Omega}) = \mu(\lambda)$ are
the Poisson parameters for sub-event $A$ and
$B$ respectively. $\mathscr{N}(k)=1/\mathscr{P}_{A}(k)$ is the normalization
factor. Here we have assumed the independent production of A and B in each
event (thermal system).

In Statistics, it is convenient to characterize  a distribution with its
moments or cumulants (see Appendix \ref{moments} for the definitions). The
first four cumulants of $\mathscr{P}_{B|A}(q|k)$ read 
\begin{eqnarray} 
	c_{1} & = & \langle\mu\rangle\equiv \mathscr{M}(k),\label{eq:mean}\\
	c_{2} & = & \langle\mu^{2}\rangle+\langle\mu\rangle-\langle\mu\rangle^{2},\label{eq:deviation}\\
	c_{3} & = & \langle\mu^{3}\rangle+\left(1-\langle\mu\rangle\right)\left[3\langle\mu{}^{2}\rangle-2\langle\mu\rangle^{2}+\langle\mu\rangle\right],\label{eq:c3t}\\
	c_{4} & = & \langle\mu^{4}\rangle+\left(\langle\mu^{3}\rangle-3\langle\mu\rangle\langle\mu^{2}\rangle+2\langle\mu\rangle^{3}\right)\left(6-4\langle\mu\rangle\right)\nonumber \\
			&  & +\langle\mu^{2}\rangle\left(7-3\langle\mu^{2}\rangle\right)+\langle\mu\rangle-7\langle\mu\rangle^{2}+2\langle\mu\rangle^{4},
	\label{eq:c4t}
\end{eqnarray} 
where$\langle(...)\rangle\equiv\mathscr{N}(k)\int d\lambda
f(\lambda)\frac{\lambda^{k}e^{-\lambda}}{k!}(...)$.  The scale variance of
$\mathscr{P}_{B|A}(q|k)$ is 
\begin{eqnarray} 
	\omega_{B} & = & 1+\frac{\langle\left(\mu-\langle\mu\rangle\right)^{2}\rangle}{\langle\mu\rangle}\geq1.\label{eq:scaledev}
\end{eqnarray} 

In generally, if we have the distribution of $f(\lambda)$ and $u(\lambda)$,
the cumulants in Eq.(\ref{eq:mean},\ref{eq:deviation},\ref{eq:c3t},\ref{eq:c4t}) can be
obtained accordingly. Here we introduce a new approach
to calculate the higher order cumulants of $\mathscr{P}_{B|A}(q|k)$ using the
distributions $\mathscr{P}_{A}(k)$ and $\mathscr{M}(k)$ measured in
experiment\footnote{In principle, the distributions $f(\lambda)$ and
$\mu(\lambda)$ can be solved from Eq.(\ref{eq:multiplicity}) and
Eq.(\ref{eq:mean}) if we known the informations of $\mathscr{P}_{A}(k)$ and
$\mathscr{M}(k)$.}.  Using  series expansion, we have
\begin{equation} 
	\mu=\sum_{m=0}^{N}a_{m}\lambda^{m}.
\end{equation}
Therefore,
\begin{eqnarray} 
	\langle\mu^{n}\rangle & = & \sum_{m_{1}=0}^{N}\sum_{m_{2}=0}^{N}..\sum_{m_{n}=0}^{N}a_{m_{1}}a_{m_{2}}...a_{m_{n}}\nonumber \\
								 &  & \times\frac{(k+\sum_{i=1}^{n}m_{i})!}{k!}\frac{\mathscr{P}_{A}(k+\sum_{i=1}^{n}m_{i})}{\mathscr{P}_{A}(k)}.
	\label{eq:ulambdan}
\end{eqnarray}
The coefficients $a_{m}$ can be extracted by fitting the
data of $\mathscr{M}(k)$ 
\begin{equation}
	\mathscr{M}(k)=\sum_{m=0}^{N}a_{m}\frac{(k+m)!}{k!}\frac{\mathscr{P}_{A}(k+m)}{\mathscr{P}_{A}(k)}.
	\label{eq:input}
\end{equation}
with a finite truncation order $N$.

Consequently, with the help of the data of 
$\mathscr{P}_{A}(k)$ and  $\mathscr{M}(k)$, 
Eq.(\ref{eq:ulambdan},\ref{eq:input}) and 
Eq.(\ref{eq:deviation},\ref{eq:c3t},\ref{eq:c4t}) provide a new approach to
calculate the second, third and fourth order cumulants of 
$\mathscr{P}_{B|A}(q|k)$. Here we have assumed the contribution from 
critical fluctuations, if any, can be neglected for the measured 
$\mathscr{P}_{A}(k)$ and $\mathscr{M}(k)$. The higher order cumulants can be
calculated analogously. 

\subsection{Net-conserved charges}
\label{sec:netcharge} 
If we
assume the independent production of positive and negative conserved charges
in each event, under the Poisson approximation, the conditional probability distribution 
of net-conserved charges can be obtained from Eq.(\ref{eq:generaleq}) as
\begin{equation} 
	\mathscr{P}_{B|A}(n|k)=\mathscr{N}(k)\int d\lambda f(\lambda)(\frac{\lambda^{k}e^{-\lambda}}{k!})\mathrm{Sk}(n;q,\lambda).
	\label{eq:netcharge}
\end{equation}
Here
$\mathrm{Sk}(n;q,\lambda)=(\mu_{+}/\mu_{-})^{n/2}I_{n}(2\sqrt{\mu_{+}\mu_{-}})\exp[-(\mu_{+}+\mu_{-})]$
is the Skellam distribution \cite{BraunMunzinger:2011dn,Aggarwal:2010wy} with Poisson
parameters $\mu_{+}=\mu_{+}(\lambda)$ and $\mu_{-}=\mu_{-}(\lambda)$ of
positive and negative-conserved charges, respectively. $n$ is the multiplicity of 
net-conserved charges in sub-event $B$. The corresponding cumulants read
\begin{eqnarray} 
	c_{2}^{N} & = & c_{2}^{\mu_{+}}+c_{2}^{\mu_{-}}-2(\langle\mu_{+}\mu_{-}\rangle-\langle\mu_{+}\rangle\langle\mu_{-}\rangle),\label{eq:idp}\\
	c_{n+1}^{N} & = & m_{n+1}^{N}-\sum_{s=0}^{n-1}\frac{n!}{s!(n-s)!}m_{n-s}^{N}c_{s+1}^{N},
\end{eqnarray} 
where $c_{n}^{\mu_{+}}$, $c_{n}^{\mu_{-}}$ are the cumulants of
positive and negative-conserved charges respectively. $m_{n}^{N}$ are the raw
moments of $\mathscr{P}_{B|A}(n|k)$. Here we give the first four moments which
will be used in the following discussions,
\begin{eqnarray} 
	m_{1}^{N} & = & \langle\mu_{+}\rangle-\langle\mu_{-}\rangle,\\ 
	m_{2}^{N} & = & \langle(\mu_{+}-\mu_{-})^{2}\rangle+\langle\mu_{+}\rangle+\langle\mu_{-}\rangle,\\
	m_{3}^{N} & = & \langle(\mu_{+}-\mu_{-})^{3}\rangle+3\langle\mu_{+}^{2}\rangle-3\langle\mu_{-}^{2}\rangle+m_{1}^{N},\\
	m_{4}^{N} & = & \langle(\mu_{+}-\mu_{-})^{4}\rangle+6\langle(\mu_{+}-\mu_{-})^{2}(\mu_{+}+\mu_{-})\rangle\nonumber \\
				 &  & +6\langle\mu_{+}^{2}+\mu_{-}^{2}\rangle+m_{2}^{N}, 
\end{eqnarray}
and
\begin{eqnarray} 
	\langle\mu_{+}^{m}\mu_{-}^{n}\rangle & = & \sum_{s_{1}=0}^{N}..\sum_{s_{m}=0}^{N}\sum_{r_{1}=0}^{N}..\sum_{r_{n}=0}^{N}a_{s_{1}}..a_{s_{m}}\nonumber \\ 
													 &  & \times\bar{a}_{r_{1}}..\bar{a}_{r_{n}}\frac{(k+\sum_{i=1}^{m}s_{i}+\sum_{i=1}^{n}r_{i})!}{k!}\nonumber \\
													 &  & \times\frac{\mathscr{P}_{A}(k+\sum_{i=1}^{m}s_{i}+\sum_{i=1}^{n}r_{i})}{\mathscr{P}_{A}(k)}.
\end{eqnarray} 
The coefficients $a_{s}$ and $\bar{a}_{r}$ are
determined by Eq.(\ref{eq:input}) with the mean value distribution of
positive and negative-conserved charges measured in experiment. Although they 
were assumed to be produced independently in each event, the relations
$c_{n}^{N}=c_{n}^{\mu_{+}}+(-1)^{n}c_{n}^{\mu_{-}}$ are broken in
event-by-event analysis (see e.g. Eq.(\ref{eq:idp})), due to the correlations of positive and 
negative-conserved charges from the distribution of PTVs.

Obviously, the statistical expectations of multiplicity distribution depend on
the multiplicity of reference particles. However, this feature has not
been taken seriously in previous studies, and only few observations have been
reported. In the following subsection, with the insufficient data, we calculate the 
approximate solutions of these high cumulants. 
We will show that these solutions can qualitatively or
quantitatively describe most of the observables related to multiplicity
fluctuations.

\subsection{Approximate solutions} 
\label{sec:approx} 
\begin{figure} 
	\begin{center} 
		\includegraphics[scale=0.4]{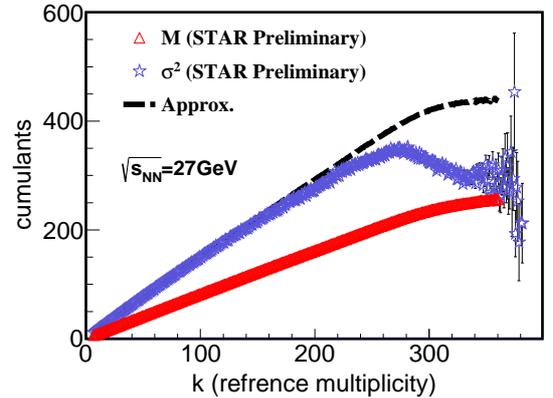}
	\end{center} 
	\caption{(Color online). Approximate solutions of $\sigma^2$
	($c_{2}$) of the total charge multiplicity distribution in Au+Au collisions
	at $\sqrt{s_{NN}}=27$GeV. The approximate solutions are obtained from
	Eq.(\ref{eq:sigma2total}). The input distribution $\mathscr{M}(k)$ are
	taken from~\cite{Tang:2014ama}.
	\label{fig:totalcharge}} 
\end{figure}

To give the analytic
solutions, we consider only the effect from distribution of volume. Due to $\mu$ and $\lambda$ 
are both proportional to volume in statistical model, the Poisson parameter $\mu$ can be written
as $\mu=b\lambda$ and $b$ is independent of $\lambda$. This consideration is also
inspired by the near-linear feature of mean value distribution $\mathscr{M}(k)$
measured in experiments (see  e.g. Fig.~\ref{fig:totalcharge}).  Secondly, except the
rapid decreasing of $\mathscr{P}_{A}(k)$ in most-central and most-peripheral
collision range, the assumption of
$\mathscr{P}_{A}(k+m)/\mathscr{P}_{A}(k)\simeq1$ is comfortable when $m$ is
not too large~\cite{Adamczyk:2012ku}. 

In general, the high order cumulants of $\mathscr{P}_{B|A}(q|k)$ and
$\mathscr{P}_{B|A}(n|k)$ in semi-central and semi-peripheral collision range
can be well described by the approximate solutions. But for the central and
peripheral collision range, the approximate solutions are questionable due to the fact that
the assumption of $\mathscr{P}_{A}(k+m)/\mathscr{P}_{A}(k)\simeq1$ becomes invalid~\cite{Xu:2016qzd}.

The approximate solutions of higher order cumulants of $\mathscr{P}_{B|A}(q|k))$ from Eq.(\ref{eq:deviation},\ref{eq:c3t},\ref{eq:c4t}) 
read
\begin{eqnarray} 
	c_{2} & = & \frac{M^{2}}{k+1}+M,\label{eq:sigma2total}\\ 
	c_{3} & = & \frac{2M^{3}}{(k+1)^{2}}+\frac{3M^{2}}{k+1}+M,\\ 
	c_{4} & = & \frac{6M^{4}}{(k+1)^{3}}+\frac{12M^{3}}{(k+1)^{2}}+\frac{7M^{2}}{k+1}+M.
\end{eqnarray} 
where $M\equiv\mathscr{M}(k)$. We find that these approximate
solutions obey the
standard NBD expectations and the NBD parameters $r$ and $p$
(Eq.(\ref{eq:NBD})) are
\begin{eqnarray}
	r &=& k+1, \\ 
	p &=& \frac{M}{M+k+1}. 
\end{eqnarray} 
The scale variance $\omega=1+M/(k+1)$
increases with $M$ while $r$ is independent of
$M$, these features have been observed in Ref.~\cite{Adare:2008ns}. In that paper, 
the authors  found
that $\omega$ increases with transverse momentum ($p_{T}$) range
of moment-analysis particles (see Fig.6 and Fig.7 in  
that paper), but $r$ (denoted as $k_{\mathrm{NBD}}$ in the reference)
show no significant $p_{T}$-dependence (see Fig.8 and Fig.9 in
that paper). This is because in Ref.~\cite{Adare:2008ns} a
narrower $p_{T}$ range correspond to a smaller $M$.  

In Fig.~\ref{fig:totalcharge}, we show the approximate solutions of
$\sigma^{2}$ of the total charge multiplicity distribution in Au+Au collisions
at $\sqrt{s_{NN}}=27$GeV as function of reference multiplicity $k$. The input
distribution $\mathscr{M}(k)$ (open triangle symbol) are taken from~\cite{Tang:2014ama}.
We find that the approximate solution (black-dashed line)
can reproduce the experimental results(open star symbol) expect
the central collision range. The deviations in most central collision are due
to the non-trivial features of $\mathscr{P}_{A}(k)$ in this range,
that make the second assumption $\mathscr{P}_{A}(k+m)/\mathscr{P}_{A}(k)\simeq1$
becomes invalid.

From the approximate solutions, we obtain the relationship among the scale variance of 
total charge hadrons $\omega_{ch}$,  positive hadrons $\omega_{+}$ and negative hadrons $\omega_{-}$
\begin{equation}
\omega_{ch}=\omega_{+} + \omega_{-} - 1.
\end{equation}
Within the accuracy errors this relations can be used to 
explain the experiment measurements from NA49 collaborations~\cite{Alt:2006jr} and PHENIX collaborations ~\cite{Adare:2008ns} 
surprisingly well, even the effect of resonance decays have not been included in the present study. Moreover, the $M/k$ ratios 
help to explain the differences on scale variance of total charge distributions measured in different centralities 
and different experiments~\cite{Alt:2006jr,Adare:2008ns,Tang:2014ama,Mukherjee:2016hrj}.

\subsubsection{Net-conserved charges}
Analogously, we obtain the approximate solutions of first four
cumulants of $\mathscr{P}_{B|A}(n|k)$ as
\begin{eqnarray} 
	c_{1}^{N} & = & M_{+}-M_{-},\\
	c_{2}^{N} & = & \frac{(M_{+}-M_{-})^{2}}{k+1}+M_{+}+M_{-},\\
	c_{3}^{N} & = & \frac{2(M_{+}-M_{-})^{3}}{(k+1)^{2}}+\frac{3(M_{+}^{2}-M_{-}^{2})}{k+1}+c_{1}^{N},\\
	c_{4}^{N} & = & \frac{6(M_{+}-M_{-})^{4}}{(k+1)^{3}}+\frac{12(M_{+}-M_{-})^{2}(M_{+}+M_{-})}{(k+1)^{2}}\nonumber \\
				 &  & +\frac{6(M_{+}^{2}+M_{-}^{2})}{k+1}+c_{2}^{N}, 
\end{eqnarray} 
where $M_{+}$ and $M_{-}$ are the mean values of positive and negative conserved
charges in a given reference multiplicity bin $k$.

\begin{figure*} 
	\begin{center} 
		\includegraphics[scale=0.4]{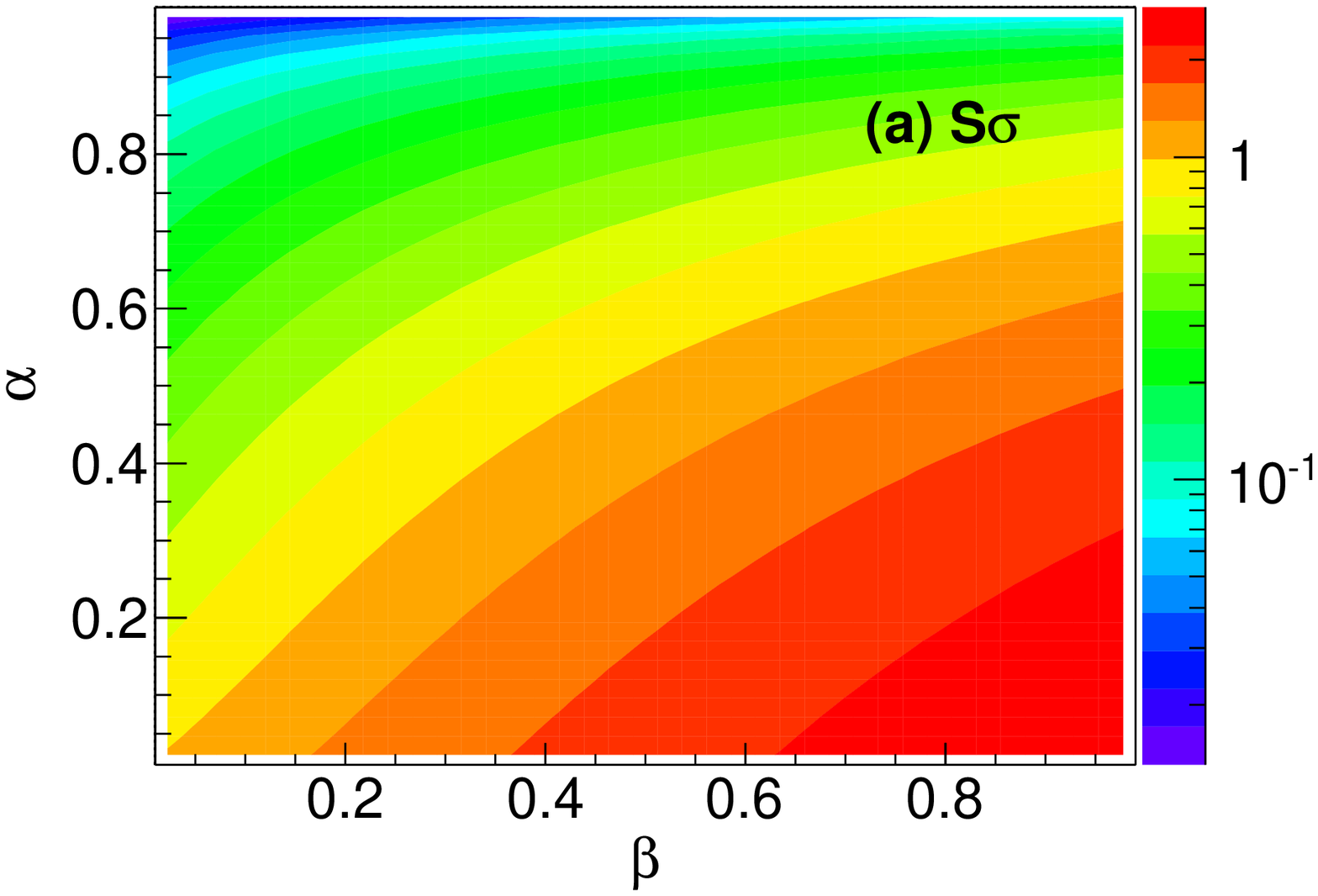}
		\includegraphics[scale=0.4]{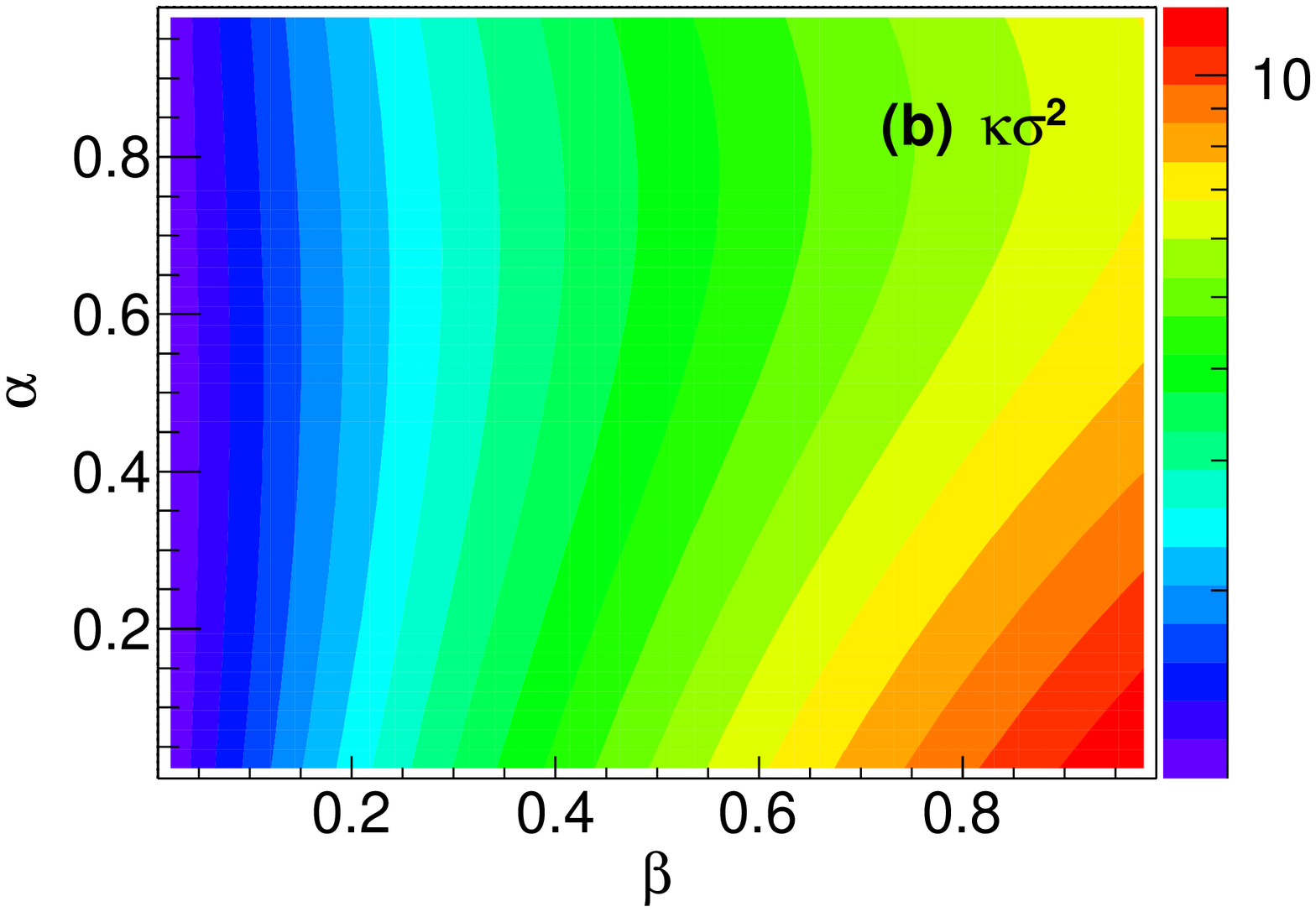}
	\end{center}
	\caption{(Color online). The $\beta-\alpha$ plane of $S\sigma$(upper panel)
		and $\kappa\sigma^{2}$(lower panel) of multiplicity distribution of
		net-conserved charges. $\beta$ is the multiplicity ratio between
		positive-conserved charges and the reference multiplicity, $\alpha$ is
	the multiplicity ratio between negative and positive-conserved charges. For
the details, see Eq.(\ref{eq:Ssigma}) and Eq.(\ref{eq:ksigma2}).
\label{fig:analysisical}} 
\end{figure*}

Due to less sensitive to the interaction volume and experimental efficiency~\cite{Karsch:2010ck,Alba:2014eba,Fu:2013gga,Aggarwal:2010wy,Adamczyk:2013dal,Adamczyk:2014fia},
the moment products $S\sigma\equiv c_{3}^{N}/c_{2}^{N}$ and
$\kappa\sigma^{2}\equiv c_{4}^{N}/c_{2}^{N}$ have been frequently discussed in
both theory and experiment. From the above
approximate solutions, we have 
\begin{eqnarray} 
	S\sigma & = & 2\beta(1-\alpha)+\frac{\beta(1-\alpha^{2})+1-\alpha}{\beta(1-\alpha)^{2}+1+\alpha},\label{eq:Ssigma}\\
	\kappa\sigma^{2} & = & 6\beta(\gamma-\frac{2\alpha}{\gamma})+1,
\label{eq:ksigma2}
\end{eqnarray}
where $\alpha=M_{-}/M_{+}$, $\beta=M_{+}/(k+1)$ and
$\gamma=\beta(1-\alpha)^{2}+1+\alpha$. If $\beta\rightarrow 0$,
Eq.(\ref{eq:Ssigma}) and Eq.(\ref{eq:ksigma2}) will back to the Skellam
expectations: $S\sigma=(1-\alpha)/(1+\alpha)$ and $\kappa\sigma^{2}=1$.

\begin{figure}
	\begin{center} 
	\includegraphics[scale=0.4]{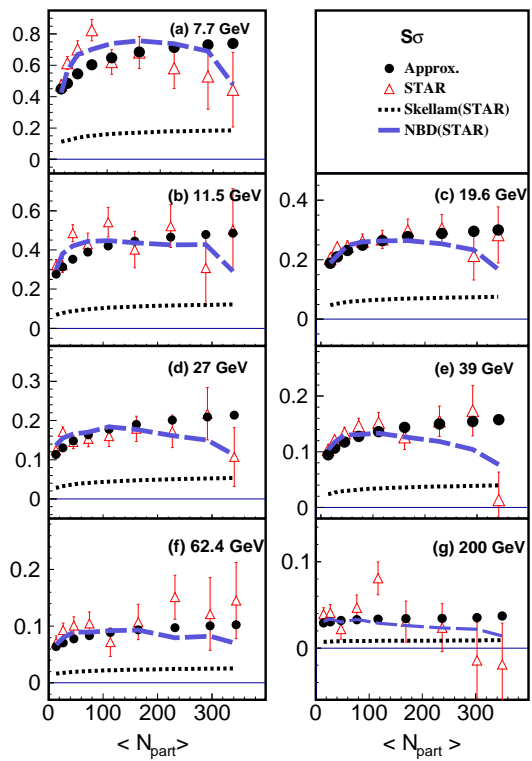} \end{center}
	\caption{(Color online). Approximate solutions of $S\sigma$ of the
			net-charge multiplicity distribution in Au+Au collisions at
			$\sqrt{s_{NN}}=7.7$ to $200$ GeV.  The data, Skellam and NBD
	baselines are taken from ~\cite{Adamczyk:2014fia}.  The approximate
	solutions are four times of the Skellam measures, see
	Eq.(\ref{eq:Ssigmanet}). 
	\label{fig:SsigmaBES}} 
\end{figure}

\begin{figure} 
	\begin{center} 
		\includegraphics[scale=0.4]{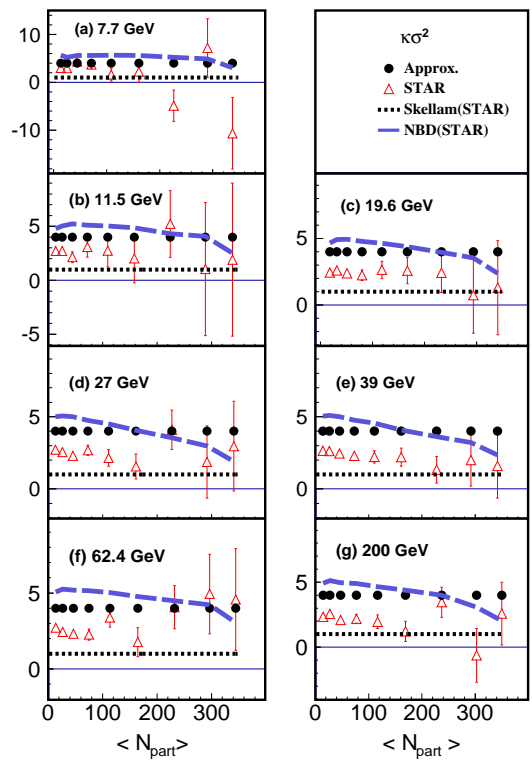}
\end{center}
	\caption{(Color online). Similar to Fig.~\ref{fig:SsigmaBES}, but for
$\kappa\sigma^2$. The approximate solutions are four times of the Skellam
measures, see Eq.(\ref{eq:ksigma2net}).  
\label{fig:ksigma2BES}}
\end{figure}

In Fig.~\ref{fig:analysisical} we show the $\beta-\alpha$ plane of $S\sigma$
and $\kappa\sigma^{2}$ 
from Eq.(\ref{eq:Ssigma}) and Eq.(\ref{eq:ksigma2}). The approximate
solutions can explain many
observations on multiplicity fluctuations except the
most-central and most-peripheral centralities:

\begin{enumerate}
 \item \textbf{Centrality resolution effect}. The moments and its products
$S\sigma$ and $\kappa\sigma^{2}$ not only dependent on the multiplicity ratio
between negative and positive conserved charges, but also depend on the
multiplicity used for centrality definition. This property has been found in
both experimental measurements and some model calculations~\cite{Luo:2013bmi},
which was considered as centrality resolution effect. More specifically, a larger
pseudorapidity range of reference multiplicity contribute to a smaller values
of $S\sigma$ and $\kappa\sigma^{2}$ due to its smaller $\beta$, and vice
versa. 

 \item \textbf{Net-charge versus net-kaon}. 
Comparison with the cumulants of
net-charges and net-kaons distributions, the $\kappa\sigma^{2}$
of net-charges distributions will be larger than the net-kaons
one due to its larger $\beta$ and $\alpha$, see Fig.~\ref{fig:analysisical}(b).
But for $S\sigma$, there is a competition between $\beta$ and $\alpha$,
because $S\sigma$ increase with $\beta$ and decrease with $\alpha$, as
it was shown in Fig.~\ref{fig:analysisical}(a). Meanwhile, due to the
smaller $\beta$ in net-kaons case, its cumulants will be more closer to the
Skellam baseline measure than in the net-charges case. These features are in consist
with data~\cite{Adamczyk:2014fia,Sarkar:2014fja}.

\item \textbf{Independent production approximation}. 
As we have mentioned
before, the independent production relations of positive and
negative-conserved charges has been violated in event-by-event analysis.
Moreover, the NBD baselines obtained by
$c_{n}^{N}=c_{n}^{\mu_{+}}+(-1)^{n}c_{n}^{\mu_{-}}$ overestimate the higher order
cumulants of net-conserved charges distributions~\cite{Adamczyk:2014fia}. However, the
corrections for $S\sigma$ and $\kappa\sigma^{2}$ depended on the parameters
$\beta$ and $\alpha$. 

\item \textbf{Quantitative estimation}. Using $(M_{+}+M_{-})\simeq k \gg
(M_{+}-M_{-})$ in the net-charge case \cite{Adamczyk:2014fia}, we have 
$\beta\simeq1/(1+\alpha)$, $\alpha\simeq1$ and 
\begin{eqnarray}
	S\sigma &\simeq& \frac{4(1-\alpha)}{1+\alpha}, \label{eq:Ssigmanet}\\
	\kappa\sigma^{2} &\simeq& 4 \label{eq:ksigma2net},
\end{eqnarray}
which are about four times of the Skellam expectations.  The results are shown in
Fig.~\ref{fig:SsigmaBES} and Fig.~\ref{fig:ksigma2BES}.  We find that the
approximate solutions of $S\sigma$ are closer to the experiment data/NBD baselines than
the Skellam baselines given in~\cite{Adamczyk:2014fia}. The
approximate solutions of $\kappa\sigma^{2}$ are colser to the NBD baselines, but fail 
to quantitatively reproduce the data. This indicate the existence of correlations 
of positive and negative charges~\cite{Adamczyk:2014fia} and/or the correlations
between the moment-analysis parameters and the reference particles. Notice that,
though it have been shown in the figures, the approximate solutions in
$0-5\%$ and $60-80\%$ centrality bins are questionable due to the non-trivial features of
$\mathscr{P}_{A}(k)$ in these ranges. 
\end{enumerate}


\subsection{Comments and discussion} \label{sec:discussion} 

In this section we have calculated the improved baseline measure of higher 
order cumulants of multiplicity distribution.  
We found that, even uner Poisson approximation, the statistical
baseline measure deviates from the Poisson measure. However, as we have 
mentioned, even in a SSE there are some other effects that make the multiplicity fluctuation
deviates from Poisson distribution. These corrections should be taken into
account especially in the case of $\beta\rightarrow 0$ when the former
deviations are small. 

In general, the two sub-events used for centrality definition and
for moment-analysis are expected to be totally independent of each event. However, the
unexpected correlations between them, as well as the correlations between the
positive and negative-conserved charges in net-conserved charges case, might contaminate our discussions.

These correlations might be one of the reason why the Binomial distribution
instead of NBD have be observed in experiment~\cite{Adamczyk:2013dal} for the
protons and anti-protons distributions. We notice that the two
sub-events used for centrality definition and for moment analysis share a common
pseudorapidity range. Using a transport dynamic model~\cite{Westfall:2014fwa},
the author found that the high order cumulants of net-proton distributions are sensitive to the
definition of reference multiplicity. Meanwhile, due to the small
$\beta$ in proton and anti-proton cases, some other corrections might overcome the correction discussed in
this work, and alter the classifications of proton and anti-proton distributions.

\section{Conclusion} 
The traditional calculations of higher order cumulants of multiplicity 
distributions are incomplete due to lack of the distribution of principal
thermodynamic variables and the probability condition from reference multiplicity. 
After including the distribution of 
principal thermodynamic variables, we have derived a general expression for 
the multiplicity distribution in terms of a conditional probability with 
arbitrary statistical ensembles and distribution of thermodynamic variables. As an application, 
we have used the general formula to calculate higher order cumulants under the Poisson 
approximation.  

We found that the improved baseline measure for multiplicity distribution mimics 
the negative binomial distribution instead of Poisson one, though the Poisson distribution 
was used as input in a specific statistical ensemble. The deviation of the new baseline measure 
from the Poisson one increases with the ratio of the mean multiplicity $M$ to the 
corresponding reference multiplicity $(k+1)$. The basic statistical expectations work well in describing
the negative binomial multiplicity distribution  measured in experiments, e.g. the cumulants (cumulant products) 
for multiplicity distribution of total (net) charges.

Similar to the trivial Poisson expectations, the basic statistical expectations can be directly constructed from experiment, 
but with the data of mean multiplicity $\mathscr{M}(k)$ and distribution of reference multiplicity $\mathscr{P}_{A}(k)$  .
However, we note that currently the exact statistical measure cannot 
be fully determined because of insufficient data. 
These data are crucial for calculation of the new baseline measure especially in most central 
collision due to non-trivial feature of $\mathscr{P}_{A}(k)$ in this range. 
The measurements of these distributions are highly expected in the future 
to pin down the exact statistical measure. 

\section*{Acknowledgments} The author would like to thanks Qun Wang for a careful reading of 
the manuscript and useful comments. The author also acknowledges 
fruitful discussions with T. S. Biro, L. J. Jiang, J. X. Li, H. C. Song, N. R. Sahoo,  
and A. H. Tang. This work is supported by China Postdoctoral Science Foundation with
grant No.~2015M580908.

\appendix

\section{Moments and cumulants} 
\label{moments}
For a probability distribution
$f(x)$, the moment-generating function can be written as,
\begin{equation}
	M(t) = \int_{-\infty}^{\infty} f(x) e^{tx}dx. 
\end{equation}
We obtain the sereies expansion,
\begin{equation} 
	M(t) = \sum_{n=0}^{\infty}m_{n}\frac{t^{n}}{n!},
\end{equation} 
where $m_{n}$ is the $n$th-order raw moment for $f(x)$
\begin{equation}
	m_{n} = \int_{-\infty}^{\infty}dx x^{n}f(x). 
\end{equation}

The cumulant-generating function is defined as 
\begin{equation}
	K(t)=\ln M(t) = \sum_{n=1}^{\infty} c_{n}\frac{t^{n}}{n!}, 
\end{equation}
where $c_{n}$ is the $n$th-order cumulant of $f(x)$. Then we have
\begin{equation}
	M(t)=\sum_{n=0}^{\infty} m_{n}\frac{t^{n}}{n!} = \exp(\sum_{n=1}^{\infty} c_{n}\frac{t^{n}}{n!}).  
\end{equation}
By taking $n$th order derivatives at $t=0$, we have 
\begin{eqnarray} 
	m_{n+1} &=& \sum_{p=0}^{n}\frac{n!}{p!(n-p)!}m_{n-p}c_{p+1},\\
	c_{n+1} &=& m_{n+1} -\sum_{p=0}^{n-1}\frac{n!}{p!(n-p)!}m_{n-p}c_{p+1},
\end{eqnarray}

The first four order explicit relation, which was frequently used in this paper,
reads,
\begin{eqnarray} 
	m_{1} &=& c_{1},\\ 
	m_{2} &=& c_{2}+c_{1}^{2},\\
	m_{3} &=& c{3}+3c_{1}c_{2}+c_{1}^{3},\\
	m_{4} &=& c_{4}+4c_{3}c_{1}+3c_{2}^{2}+6c_{2}c_{1}^{2}+c_{1}^{4},
\end{eqnarray} 
and 
\begin{eqnarray}
	c_{1} &=& m_{1}\equiv\mu,\\ 
	c_{2} &=& m_{2}-m_{1}^{2}\equiv\sigma^{2},\\
	c_{3} &=& m_{3}-3m_{2}m_{1}+2m_{1}^{3}\equiv S\sigma^{3},\\ 
	c_{4} &=& m_{4}-4m_{3}c_{1}-3m_{2}^{2}+12m_{2}m_{1}^{2}-6m_{1}^{4}\nonumber \\
			&\equiv&\kappa\sigma^{4},
\end{eqnarray}
where $\mu$, $\sigma^{2}$, $S$ and $\kappa$ are
mean value, variance, skewness and kurtosis of
probability distribution $f(x)$, respectively.

For the Poisson distribution, we have
\begin{equation}
	c_{1}=c_{2}=c_{3}=c_{4}=\lambda,
\end{equation} 
where $\lambda$ is the Poisson parameter shown in Eq.(\ref{eq:multiplicity0}). The scale variance for Poisson distribution is $\omega=c_{2}/c_{1}=1$. 

For the NBD, we have
\begin{eqnarray}
	c_{1} &=& \frac{rp}{1-p},\\ 
	c_{2} &=& \frac{rp}{(1-p)^{2}},\\
	c_{3} &=& \frac{rp(1+p)}{(1-p)^{3}},\\ 
	c_{4} &=& \frac{6rp^{2}}{(1-p)^{4}} + \frac{rp}{(1-p)^{2}}, 
\end{eqnarray} 
where $r$ and $p$ are NBD parameters shown in Eq.(\ref{eq:NBD}). The scale variance for NBD is $\omega=c_{2}/c_{1}=1/(1-p)>1$.

\bibliography{ref}

\end{document}